\documentclass{llncs}
\usepackage{makeidx}  
\usepackage[ruled,vlined,lined,boxed,commentsnumbered]{algorithm2e}
\usepackage{cite}
\usepackage{caption}
\usepackage[usenames,dvipsnames,svgnames,table]{xcolor}
\usepackage[titletoc]{appendix}
\usepackage{graphicx}
\usepackage{wrapfig, blindtext}
\usepackage{amssymb, amsmath} 
\usepackage{mathtools}
\usepackage[margin=1.6in]{geometry}

\newcommand{\bigO}{\mathcal{O}}

\newcommand{\llen}{\ell}
\newcommand{\xlen}{m}
\newcommand{\qbl}{n_q}

\newcommand{\rk}{\phi_q}

\renewcommand\appendix{\par
	\setcounter{section}{0}
	\setcounter{subsubsection}{0}
	\setcounter{figure}{0}
	\setcounter{table}{0}
	\renewcommand\thesection{Appendix \Alph{section}}
	\renewcommand\thefigure{\Alph{section}\arabic{figure}}
	\renewcommand\thetable{\Alph{section}\arabic{table}}
}

\begin{document}
	
	\title{Fast Longest Common Extensions in Small Space}
	
	\author{Alberto Policriti\inst{1,2} \and Nicola Prezza\inst{1}}
	
	\institute{University of Udine, Department of Informatics, Mathematics, and Physics, Italy \and Applied Genomics Institute, Udine}
	
	\maketitle
	
	\begin{abstract}
		In this paper we address the  longest common extension (LCE) problem: to compute the length $\llen$ of the longest common prefix between any two suffixes of $T\in \Sigma^n$ with $ \Sigma = \{0, \ldots \sigma-1\} $.  
		We present two fast and space-efficient solutions based on (Karp-Rabin) \textit{fingerprinting} and \textit{sampling}. Our first data structure exploits properties of Mersenne prime numbers when used as moduli of the Karp-Rabin hash function and takes $n\lceil \log_2\sigma\rceil$ bits of space. Our second structure works with any prime modulus and takes $n\lceil \log_2\sigma\rceil + n/w + w\log_2 n$ bits of space ($ w $ memory-word size). Both structures support 
		$\bigO\left(\xlen\log\sigma/w \right)$-time extraction of any length-$\xlen$ text substring, $\bigO(\log\llen)$-time LCE queries with high probability, and can be built in optimal $\bigO(n)$ time. In the first case, ours is the first result showing that it is possible to answer LCE queries in $o(n)$ time while using only $\bigO(1)$ words on top of the space required to store the text. Our results improve the state of the art in space usage, query times, and preprocessing times and are extremely practical: we present a C++ implementation that is very fast and space-efficient in practice.
		
	\end{abstract}
	
	\section{Introduction}
	Computing using a limited amount of space is a difficult and intriguing problem. In text-processing, \textit{fingerprinting} and \textit{sampling} are two natural and effective ways to tackle the problem. In this paper we use such techniques to improve on state-of-the-art solutions to the \textit{longest common extension} problem, using  properties of Mersenne prime numbers used as moduli in fingerprint computation.
	Let $T\in\Sigma^n$ be a length-$n$ text over an alphabet $\Sigma=\{0,\dots,\sigma-1\}$. The longest common extension (LCE) problem asks to build a data structure on $T$ supporting fast LCE queries: given $0\leq i,j<n$, return the length $\llen=LCE(i,j)$ of the longest common prefix of $T[i,\dots, n-1]$ and $T[j,\dots, n-1]$. This problem has several important applications in suffix-sorting, computing palindrome factors~\cite{breslauer1995finding,allouche2003palindrome}, identification of repeats~\cite{landau1993algorithm,kolpakov2014searching}, and string matching\cite{myers1986ano,landau1986introducing,amir2004faster}. 
	
	In this paper, we discuss fast and  space-efficient LCE data structures. $\Theta(n\log n)$-bits solutions are known, guaranteeing constant-time LCE queries: such techniques include (i) building a $\bigO(1)$-time lowest common ancestor structure (see, e.g.,~\cite{harel1984fast}) on the suffix tree of the text (ST+LCA) and (ii) combining a longest common prefix structure, a $\bigO(1)$-time range minimum query structure (see, e.g.~\cite{fischer2006theoretical}), and the inverted suffix array (RMQ+LCP). On another extreme, one could simply store $T$ in $n\lceil\log_2\sigma\rceil$ bits and answer LCE queries in $\bigO(\llen)$ time by comparing $T[i,\dots, n-1]$ and $T[j,\dots, n-1]$ character by character. In their recent works~\cite{bille2015longest,bille2014time}, Bille et al. propose a number of solutions to this problem that trade query-time for space-usage. Their most practical solution is a Monte Carlo data structure requiring $\bigO(nw/\tau)$ bits on top of the text, $w$ and $0 <\tau\leq n$ being the memory word size and the block size, respectively, and answering LCE queries w.h.p. (with high probability) in $\bigO(\tau)$ time. This structure can be built in $\bigO(n\log(n/\tau))$ time. They also propose deterministic and Las Vegas structures with the same space-time bounds, but their preprocessing times---$\bigO(n^{2+\epsilon})$ and $\bigO(n^{3/2})$---make them not practical on big input texts. Deterministic data structures with similar time bounds but considerably improved construction times (of $\bigO(n\tau)$) have been recently described by Tanimura et al. in~\cite{tanimura2016}.	Of interest is also the recent work described in~\cite{nishimoto2015dynamic,inenaga2015faster}, where the authors propose the first compressed LCE data structure based on SLPs (straight-line programs).
	
	In this work we present two LCE data structures based on Karp-Rabin fingerprinting. The first structure takes $n\lceil\log_2\sigma\rceil$ bits of space---i.e. exactly the same space of the plain text---and replaces the text in that it supports $\bigO\left(\xlen\log\sigma/w\right)$ average-case time extraction of any length-$\xlen$ text substring. LCE queries are answered w.h.p. (with high probability) in $\bigO(\log\llen)$ average-case time. 
	In the worst case, text extraction and LCE queries are supported in $\bigO\left(\xlen\log\sigma + \log n\right)$ and $\bigO(\log\llen\cdot \log n)$ time, respectively.
	This structure exploits particular properties of Mersenne prime numbers when used as moduli of the Karp-Rabin hash function in order to achieve fast query times and the information-theoretic optimal space usage.
	
	We then extend this technique to work with general prime moduli. Our second structure takes $n\lceil\log_2\sigma\rceil + n/w +w\log_2 n$ bits of space and supports---in the worst case---optimal $\bigO\left(\xlen\log\sigma/w\right)$-time substring extraction and $\bigO(\log\llen)$-time LCE queries w.h.p. 
	
	Both structures can be built in optimal $\bigO(n)$ time. Our first result improves the state-of-the art~\cite{bille2015longest} in space, query times, and preprocessing times, but requires the modulus to be a Mersenne prime. Our second result removes this assumption and improves query times and preprocessing times of the Monte Carlo structure presented in~\cite{bille2015longest} while using its same space for $\tau=w^2$.
	
	In Table \ref{tab: state of the art} we give an overview of the main solutions for the LCE problem described in literature.
	
	\renewcommand{\arraystretch}{1.5}
	
	\begin{table}
		\vspace{.5cm}
		\centering
		\begin{tabular}{| c | c | c | c | c | c |}	\hline
			Space (bits) & Query time & correct & build time & Notes & Reference\\\hline 
			$\bigO(n\log n)$ & $\bigO(1)$ & always & $\bigO(n)$  &---& $ST + LCA$\\\hline
			$\bigO(n\log n)$ & $\bigO(1)$ & always & $\bigO(n)$ &---& $RMQ + LCP$\\\hline
			$n\lceil\log_2\sigma\rceil + \bigO(nw/\tau)$ & $\bigO(\tau)$ & w.h.p. & $\bigO(n\log(n/\tau))$ &$0<\tau\leq n$& \cite{bille2015longest}\\\hline
			$n\lceil\log_2\sigma\rceil + \bigO(nw/\tau)$ & $\bigO(\tau)$ & always & $\bigO(n^{2+\epsilon})$ &$0<\tau\leq n$&\cite{bille2015longest}\\\hline
			$n\lceil\log_2\sigma\rceil + \bigO(nw/\tau)$ & $\bigO(\tau)$ & always & $\bigO(n^{3/2})$ w.h.p. &$0<\tau\leq n$&\cite{bille2015longest}\\\hline
			$n\lceil\log_2\sigma\rceil + \bigO(nw/\tau)$ & $\bigO(\tau\log\tau)$ & always & $\bigO(n\tau)$ &$0<\tau\leq n$&\cite{tanimura2016}\\\hline
			$n\lceil\log_2\sigma\rceil$ & $\bigO(\llen)$ & always & $\bigO(1)$ &---& store only $T$\\\hline
			$n\lceil\log_2\sigma\rceil$ & $\bigO(\log\llen)$ avg. & w.h.p. & $\bigO(n)$ &$q=2^p-1$& this paper\\\hline
			$n\lceil\log_2\sigma\rceil$ & $\bigO(\log\llen\cdot \log n)$ & w.h.p. & $\bigO(n)$ &$q=2^p-1$& this paper\\\hline
			$n\lceil\log_2\sigma\rceil + n/w + w\log_2 n$ & $\bigO(\log\llen)$ & w.h.p. & $\bigO(n)$ &---& this paper\\\hline
			$\bigO(zw\log n\log^*n)$ & $\bigO(\log n\log^*n)$ & always & $\bigO(n\log z)$  &---& \cite{nishimoto2015dynamic,inenaga2015faster} \\\hline
		\end{tabular}
		\vspace{.5cm}
		\caption{Main solutions for the LCE problem described in literature. Column \emph{correct} indicates whether the structures returns \emph{always} or just \emph{with high probability} (i.e. probability at least $1-2^{-w}$) the correct LCE. We account also for the space to store the plain text if this is required to answer LCE queries. For clarity we omit the $\bigO(w)$ space term that is always present in all solutions (note that even to store the plain text in practice we need a constant number of pointers to memory). $\llen = LCE(i,j)$ is the output of the LCE query. $q$ is the prime modulus used in the Karp-Rabin hash function, and $q=2^p-1$ means that $q$ must be a Mersenne prime. The  ``avg.'' after the query time (second column) means that times are on average on uniformly distributed texts. If not specified, times are worst-case. $z$ is the number of phrases of the LZ77 parsing without self-references. $\log^*n$ is the iterated logarithm function.}\label{tab: state of the art}
	\end{table}

	\section{Preliminaries}
	
	We assume that our input text is drawn from an integer alphabet: $T\in\{0,\dots,\sigma-1\}^n$. Throughout the description of our strategy we assume without loss of generality that $\sigma = 2$. If this is not the case, then we can simply build a binary text $T'\in\{0,1\}^{n\cdot b}$, where $b = \lceil\log_2\sigma\rceil$, by concatenating the numbers $T[0], T[1], \dots, T[n-1]$ written in binary and answer LCE queries on $T$ as $T.LCE(i,j) = \lfloor T'.LCE(i\cdot b, j\cdot b)/b\rfloor$. 
	
	Let $w$ be the memory word size (in bits). We assume, as usual, that $\log n\in\bigO(w)$. For the above considerations, in our final results the only constraint we will impose on the alphabet size $\sigma$ is that $\log \sigma\in\bigO(w)$, so that we can manipulate alphabet characters in constant time. 
	
	\subsection{Karp-Rabin Fingerprinting}
	
	We adopt the fingerprinting approach of Karp and Rabin~\cite{karp1987efficient} to efficiently compare text substrings. The idea behind Karp-Rabin's technique is to represent a string $S\in\Sigma^n$ as a number in base $\sigma$ modulo a prime $q$. Their hash function is defined as:
	$$
	\rk(S) = \sum_{i=0}^{n-1} S[i]\sigma^{n-i-1}\ \mod q
	$$
	Under the assumption that $q$ fits in a constant number of memory words, this function can be used to test for string equality in constant time. Although there exists the possibility of incurring into false positives---i.e. $\rk(S)=\rk(S')$ but $S\neq S'$---, in~\cite{karp1987efficient} the authors show that, if $q$ is a prime chosen uniformly at random, then this probability is only $1/q$. 
	
	Function $\rk$ has several useful algebraic properties. In our case, we will exploit the fact that $\rk(T)$ can be obtained from $\rk(ST)$ and $\rk(S)$ with just a multiplication and a subtraction modulo $q$ (more details in the next sections).
	
	To simplify notation, for any $i>j\geq 0$ we define $\rk(T[i,\dots, j]) = \rk(\epsilon) = 0$ ($\epsilon$ is the empty string).
	
	
	\subsection{Mersenne Primes}\label{sec: Mersenne}
	
	The key point in our first result is the use of Mersenne prime numbers---i.e. primes of the form $q = 2^p-1$---as modulus of the Karp-Rabin function $\rk$. First of all, note that $2^p-1$ is a $p$-bits integer and its binary representation is composed exclusively of 1's. This fact implies that \emph{any} combination of $p$ bits except $q$ itself is smaller than the modulus $q$, and will be crucial in order to obtain the information-theoretic optimal space usage in our structure. The second property we use is related to the computation of exponentials modulo $q$. With generic $q$, the computation of $x\cdot2^e\mod q$, with $x<q$ and $e\in\mathbb{N}$, requires an exponentiation---$\bigO(\log e)$ steps with the fast exponentiation algorithm---and a constant-time multiplication. Here we prove that this operation can be implemented in constant time if $q$ is a Mersenne prime. Let $q=2^p-1$. We have that
	$$
	\begin{array}{ll}
	2^e  &  = 2^{p\cdot \lfloor e/p \rfloor + e\ mod\ p}\\
	& = 2^{p\cdot \lfloor e/p \rfloor}\cdot 2^{e\ mod\ p}\\
	& = (2^p)^{\lfloor e/p \rfloor}\cdot 2^{e\ mod\ p}\\
	& \equiv_q 1^{\lfloor e/p \rfloor}\cdot 2^{e\ mod\ p}\\
	& \equiv_q 2^{e\ mod\ p}\\
	\end{array}
	$$
	Now, let $x_0x_1\dots x_{p-1}$, with $x_i\in\{0,1\}$, be the binary representation of $x<q$: $x = \sum_{i=0}^{p-1} x_i\cdot 2^{p-i-1}$. Then:
	$$
	\begin{array}{ll}
	x\cdot 2^e  &  = (\sum_{i=0}^{p-1} x_i\cdot 2^{p-i-1})\cdot 2^e\\
	& = \sum_{i=0}^{p-1} x_i\cdot 2^{p-i-1+e}\\
	& \equiv_q \sum_{i=0}^{p-1} x_i\cdot 2^{(p-i-1+e)\ mod\ p}
	\end{array}
	$$
	The above equality means that $x\cdot 2^e\mod (2^p-1)$ can be computed by left-rotating the binary representation of $x$ by $e\mod p$ positions. This can be done in $\bigO(1)$ time with a constant number of bitwise shift, OR, and mask operations.
	
	Our first structure will make use of Mersenne primes as moduli of the Karp-Rabin hash function. Note that the number of known Mersenne primes is too small to use them as source of randomness in a randomized algorithm (there are only 9 Mersenne primes fitting in a 64-bits memory word), so our first structure will return the correct LCE with high probability on uniform inputs only. We will remove this assumption in our second structure, which will work with any prime modulus. 
	
	\section{A In-Place Monte Carlo LCE structure}
	
	Our approach follows the same general idea described in~\cite{bille2015longest}: we divide the text in blocks of $\tau>0$ bits and we sample by computing the Karp-Rabin fingerprint of text prefixes ending at block boundaries. This structure permits to test (with high probability) the equality of any two text substrings and can therefore be used to compute the LCE of two text suffixes with a binary search. The key difference with the work in~\cite{bille2015longest} is that we show how this sampling of fingerprints can \emph{replace} the text while taking exactly its size ($n$ bits): we achieve this by carefully choosing $\tau$ and the prime modulus $q$ so that each block can be decoded unambiguously. In our first solution, we use primes of the form $q = 2^\tau-1$ (Mersenne primes). With this choice we can decode unambiguously all combinations of $\tau$ bits except the one corresponding to the value $q$ (which is congruent to $0$ modulo $q$). Blocks storing the value $q$ can however be discarded and replaced with an array containing their positions inside the array of blocks. In this way, we do not waste any additional bit other than the $n$ bits required to store the text and we are able to reconstruct it. Finally, since $x\cdot 2^e \mod (2^\tau-1)$ can be computed in constant time for any $x,e\in\mathbb{N}$, we do not need to sample \textit{values} $2^i\mod q$ as done in~\cite{bille2015longest} in order to speed-up fingerprint computations.
	
	\subsection{Fingerprint Sampling}\label{sec:fingerprint sampling}
	
	Let $w$ be the memory word size, $\tau\in\bigO(w)$ the block size and $q = 2^\tau-1$ the Mersenne modulus. Without loss of generality, we assume that $n$ is a multiple of $\tau$ (the general case can be reduced to this case by left-padding the text with $\tau-(n\mod\tau)$ bits). 
	

	Let $B, P'\in [0, q-1]^{n/\tau}$ be the integer arrays defined as
	$$
	B[i] = \sum_{j=0}^{\tau-1} 2^{\tau-j-1} \cdot T[i\cdot\tau+j],\ \ \ i=0,\dots, n/\tau - 1
	$$
	and
	$$
	P'[i] = \sum_{j=0}^{i} 2^{(i-j)\cdot \tau}\cdot B[j] \mod q = \rk(T[0,\dots,(i+1)\cdot\tau-1])
	$$
	To simplify notation, let $P'[-1]=0$. If $B[i]<q$, then $B[i]$ can be retrieved from $P'$ as follows:
	\begin{equation}\label{eq:B}
		B[i] = P'[i] - 2^\tau\cdot P'[i-1] \mod q
	\end{equation}
	However, if $B[i]=q$ for some $i$, the above operation yields the value $0$ instead of $q$ so $P'$ alone does not replace the text. We can remove such ambiguity by marking with a bitvector $Q\in \{0,1\}^{n/\tau}$ blocks such that $B[i]=q$:
	
	\begin{equation}\label{eq:Q}
		B[i] = q \Leftrightarrow Q[i] = 1,\ \ \ i=0,\dots,n/\tau - 1
	\end{equation}
	
	Let $\qbl$ be the number of blocks such that $B[i] = q$. We store $Q$ as an increasing sequence of integers $S_Q=\langle i\ :\ Q[i]=1,\ i=0,\dots,n/\tau-1\rangle$ so that its space usage is of $\qbl\cdot\log\frac{n}{\tau}$ bits.
	
	For simplicity we assume that $B[0]\neq q$ (if this is not the case we can left-pad the text with $\tau$ zeros). At this point, we do not need to store anymore values $P'[i]$ for $i$'s such that $ B[i]=q$ as the following property holds for $i>0$:
	$$
	B[i] = q\ \ \Rightarrow\ \  P'[i] = P'[i-1]\cdot 2^\tau + B[i] \mod q = P'[i-1]\cdot 2^\tau\mod q$$
	Let $B[i] = q$ and $k$ be the smallest integer such that $B[i-k]\neq q$. By applying recursively the above equality, we get
	\begin{equation}\label{eq: get P'}
		P'[i] = P'[i-k]\cdot 2^{k\cdot \tau} \mod q
	\end{equation}

	The above considerations imply that we can replace $P'$ with a vector $P\in [0,q-1]^{n/\tau - \qbl}$ containing only the $P'[i]$'s such that $B[i]\neq q$:
	$$
	P = \langle P'[i]\ :\ B[i] \neq q,\ i=0,\dots,n/\tau-1\rangle
	$$
	Before showing how we can retrieve $P'$ values using $P$ and $Q$, we describe how to efficiently answer to the following queries on $Q$:
	
	\begin{enumerate}
		\item \texttt{Access}: return $Q[i]$
		\item \texttt{Rank}: $Q.rank_b(i),\ 0\leq i \leq n/\tau,\ b\in\{0,1\}$. Return the number of bits equal to $b$ before position $i$ excluded
		\item \texttt{0-predecessor}: $Q.pred(i)$. Return the rightmost position $j\leq i$ such that $Q[j]=0$
	\end{enumerate}
	
	\texttt{access} requires just a binary search on $S_Q$. The same holds for $Q.rank_1(i)$: return, using a binary-search on $S_Q$, the number of elements strictly smaller than $i$. At this point, $Q.rank_0(i)$ is equal to $i - Q.rank_1(i)$. To answer $Q.pred(i)$, where $Q[i]=1$ (if $Q[i]=0$ the answer is $i$), we can use again binary search on $S_Q$. The idea is that we have to find---in the bitvector $Q$---the 0 preceding the run of 1's containing $Q[i]$; in $S_Q$, this run translates to a sub-sequence of integers with the property that the difference between any two of them is equal to their distance in the array $S_Q$. This leads to the following binary search criteria. Let $k=Q.rank_1(i)$. We start the search in the interval $S_Q[l,\dots, r]$, with $l=0,\ r=k$, and we pick the middle $t$ of this interval: $t=(l+r)/2$. First, check if $r=t$ or if $S_Q[r-t]-S_Q[r-t-1]>1$: in these cases, return $S_Q[r-t]-1$. If this is not the case, if $S_Q[r] - S_Q[r-t] = t$ then recurse in $S_Q[l,\dots, r-t-1]$, otherwise in $S_Q[r-t,\dots,r]$.
	
	At this point, let $t = Q.pred(i)$. By combining Equations \ref{eq:B}, \ref{eq:Q}, and \ref{eq: get P'} we obtain the following relations for every $0\leq i < n/\tau$:
	
	\begin{equation}\label{eq: relations}
		\left\{\begin{array}{ll}
			P'[i] & = P[Q.rank_0(t)]\cdot 2^{\tau\cdot (i-t)} \mod q\\
			B[i] & = Q[i]\cdot q + (1-Q[i])(P'[i] - 2^\tau\cdot P'[i-1]\mod q)
		\end{array}\right.
	\end{equation}
	
	Since multiplication by any power of 2 modulo $q=2^\tau-1$ takes constant time, by using $P$ and $Q$ and applying Equations \ref{eq: relations} we can (remember that $\qbl=|S_Q|$):
	
	\begin{enumerate}
		\item Retrieve any $B[i]$ in $\bigO(\log \qbl)$ time
		\item Retrieve any $P'[i]$ in $\bigO(\log \qbl)$ time
	\end{enumerate}
	
	Note that query 1. can be directly used to extract any text substring of length $\xlen$ in $\bigO(\lceil \xlen/w\rceil\cdot \log \qbl) \subseteq \bigO((\xlen/w+1)\cdot \log\qbl) \subseteq \bigO(\xlen + \log n)$ time. 
	
	\subsubsection{Space usage}
	
	We add the constraint $\tau \geq \log_2 n > \log_2\frac{n}{\tau}$: a text pointer must fit in $\tau$ bits (we just have to use a big enough Mersenne prime in order to satisfy this requirement). 
	The size of our structure is:
	$$
	\begin{array}{ll}
	|P|\cdot \tau + |S_Q|\cdot \log\frac{n}{\tau} & = \left(\frac{n}{\tau}-\qbl\right)\tau + \qbl\log\frac{n}{\tau} \\
	& \leq \left(\frac{n}{\tau}-\qbl\right)\tau + \qbl\cdot \tau\\
	& = n
	\end{array}
	$$
	bits. This is exactly the space required by the naive solution that answer LCE queries by comparing text suffixes character by character.
	
	\subsubsection{Building the structure}
	
	We analyze the construction algorithm assuming a general alphabet of size $\sigma\in\bigO(2^w)$. In $\bigO(n)$ time we read the text and pack its binary representation in blocks of $\tau$ bits; this list of blocks is array $B$. Then, array $P'$ can be easily constructed from $B$ in $\bigO(n/\tau)$ time by performing $|P'|-1$ additions and multiplications by $2^\tau$ modulo $q$. Using $B$ and $P'$, we can compute $P$ and $S_Q$ in time proportional to their size ($\bigO(n/\tau)$) and then discard $B$ and $P'$. 
	
	\subsection{Answering LCE Queries}
	
	First of all, we show how to compute the fingerprint $\rk(T[0,\dots, i])$ of any text prefix. Let $j=\lfloor i/\tau\rfloor$. Then, 
	$$
	\begin{array}{ll}
	\rk(T[0,\dots, i]) & = \rk(T[0,\dots, j\cdot\tau-1])\cdot 2^{i-j\cdot\tau+1} + \rk(T[j\cdot\tau, \dots, i]) \mod q \\
	& = P'[j-1]\cdot 2^{i-j\cdot\tau+1} + \lfloor B[j]/2^{p-i+j\cdot\tau-1}\rfloor \mod q
	\end{array}
	$$
	Since multiplication by any power of 2 takes constant time, we obtain that the computation of $\rk(T[0,\dots, i])$ takes $\bigO(\log\qbl)$ time with our structure for all $0\leq i < n$.
	
	We now have to show how to compute the fingerprint $\rk(T[i,\dots, j]),\ j\geq i$ of any text substring. This can be easily achieved by means of the equality:
	
	\begin{equation}\label{eq: rk substring}
		\rk(T[i,\dots, j]) = \rk(T[0,\dots, j]) - \rk(T[0,\dots, i-1])\cdot 2^{j-i+1} \mod q
	\end{equation}
	
	Again, since multiplication by a power of 2 takes constant time, the above operation takes $\bigO(\log\qbl)$ time with our structure. At this point, we can easily answer $LCE(i,j)$ by comparing $\rk(T[i,\dots, i+k])$ with $\rk(T[j,\dots, j+k])$ for $\bigO(\log n)$ values of $k$ with binary search. We can furthermore improve query times by performing an exponential search before applying the binary search procedure. We compare $\rk(T[i,\dots, i+k])$ with $\rk(T[j,\dots, j+k])$ for $k=2^0, 2^1, 2^2, \dots$ until the two fingerprints are different. Letting $\llen = LCE(i,j)$, this procedure terminates in $\bigO(\log\llen)$ steps. We then apply the binary search procedure described above on the interval of size $\bigO(\llen)$ obtained with the exponential search. Each exponential and binary search step take $\bigO(\log\qbl)$ time.
	
	\subsubsection{Analysis}
	
	First of all, note that if $\sigma$ is not a power of 2 then the binary representation of any alphabet character contains at least one bit equal to 0. Then, by choosing $\tau\geq 2\cdot \lceil \log_2\sigma\rceil$, we get that $B[i]\neq q$ for every $0\leq i < n/\tau$ (because every block of $\tau$ bits contains at least one bit equal to 0) and therefore $\qbl=0$. We obtain (switching to a alphabet of size $\sigma\in\bigO(2^w)$):
	
	\begin{theorem}\label{th: alphabet}
		If $\sigma\neq 2^e$, $e\in\mathbb{N}$, then our data structure takes $n\lceil \log_2\sigma\rceil$ bits of space and supports extraction of any length-$\xlen$ text substring and LCE queries w.h.p. in 
		$\bigO\left(\xlen\log\sigma/w\right)$ and $\bigO(\log\llen)$ worst-case time, respectively. The structure can be built in optimal $\bigO(n)$ time. 
	\end{theorem}
	
	The above time bounds hold also in the average case for any alphabet size. Assuming a uniform distribution of the text, the probability of any block $B[i]$ having value $q$ is only $(1/2)^\tau$. Since $\tau \geq \log_2 n$, this probability is $\bigO(1/n)$. It follows that, in the average case, the number $\qbl$ of such blocks is $\bigO(1)$. We obtain:
	
	\begin{theorem}\label{th: avg}
		Our data structure takes $n\lceil \log_2\sigma\rceil$ bits of space and supports extraction of any length-$\xlen$ text substring and LCE queries w.h.p. in 
		$\bigO\left(\xlen\log\sigma/w\right)$ and $\bigO(\log\llen)$ average-case time, respectively. The structure can be built in optimal $\bigO(n)$ time. 
	\end{theorem}
	
	In the worst case and for any alphabet size, the following bounds hold:
	
	\begin{theorem}
		Our data structure takes $n\lceil \log_2\sigma\rceil$ bits of space and supports extraction of any length-$\xlen$ text substring and LCE queries w.h.p. in 
		$\bigO\left(\xlen\log\sigma + \log n\right)$ and $\bigO(\log\llen\cdot \log n)$ worst-case time, respectively. The structure can be built in optimal $\bigO(n)$ time. 
	\end{theorem}
	
	Note that in order to guarantee the time bounds of Theorem \ref{th: alphabet} in the worst case for any alphabet size, we can always add an extra character to the alphabet. However, in case $\sigma=2^e$ for some $e>0$, this comes at the price of using $n$ additional bits on top of the optimal $n\cdot e$. The solution we give in the next section works with any alphabet size and reduces this overhead to $n/w+w\log_2n$ bits, while still guaranteeing the time bounds of Theorem \ref{th: alphabet} in the worst case.

	\section{A Faster Monte Carlo LCE structure}
	
	In this section we generalize our idea to any prime $q$. This generalization comes at the price of using $n/w + w\log_2 n$ additional bits of space.
	
	Let $q$ be our prime modulus and $\tau = \lceil\log_2 q\rceil$ the number of bits required to write numbers modulo $q$. We break the text in blocks $B[0],\dots,B[n/\tau-1]$ of $\tau$ bits each as done in the previous section, and we compute array $P'$ storing the Karp-Rabin fingerprint of text prefixes ending at block boundaries. We then discard $B$ and keep only $P'$. As noted in section \ref{sec:fingerprint sampling}, $P'$ is sufficient to decode any text block $B[i]$ such that $B[i]<q$. If $B[i]\geq q$, note that 
	$$B[i]-q \leq (2^\tau-1)-2^{\tau-1} < 2^{\tau-1} \leq q$$ 
	so $B[i]\mod q = B[i]-q$ holds. It is therefore sufficient to add a bitvector $D[0,\dots,n/\tau-1]$ marking with a '0' blocks such that $B[i]<q$ and with a '1' blocks such that $B[i]\geq q$ and retrieve $B$'s values as
	\begin{equation}\label{eq:B1}
		B[i] = \left(P'[i] - 2^\tau\cdot P'[i-1] \mod q\right) + D[i]\cdot q
	\end{equation}
	Note that this is a generalization of the idea presented in the previous section, the only difference being that with Mersenne primes there is only one block value greater than or equal to $q$ (i.e. $q$ itself). In that case, this difference permits to save the $n/w$ bits of array $D$. Since we store the full $P'$, extracting each $B[i]$ takes constant time\footnote{Note: computing $2^\tau\mod q$ in Equation \ref{eq:B} takes constant time since $2^\tau$ fits in a memory word.}. Computing $LCE(i,j)$, where we assume $i\leq j$ (swapping $i$ and $j$ does not change the result), requires $\bigO(\log n)$ binary search steps; at this point, since we do not pre-compute and store (i.e. sample) values $2^{i}\mod q,\ i=0,\dots,n-1$, at each binary search step we are forced to spend $\bigO(\log n)$ additional time to compute powers of 2 modulo $q$ (with the fast exponentiation algorithm) in order to evaluate Equation \ref{eq: rk substring}. We can avoid this overhead by pre-computing and storing (in $\tau\log_2 n$ bits) values $z_i = 2^{2^i}\mod q,\ i=0,\dots,\lfloor\log_2 n\rfloor$ and by performing binary search by splitting interval lengths in correspondence of powers of 2 as follows. First, note that $z_0=2$ and $z_{i+1} = (z_i)^2\mod q$, so these values can be pre-computed in $\bigO(\log n)$ time. Let the notation $\langle i,j,e,k\rangle$, with $0\leq i,j,e,k<n$ and $e<k$, denote that we already verified that $T[i,\dots, i+e-1] = T[j,\dots, j+e-1]$ and $T[i,\dots, i+k-1] \neq T[j,\dots, j+k-1]$. We use this notation to indicate the state of a binary search step, and start from state $\langle i,j,0,n-j \rangle$ (we assume for simplicity that $T[i,\dots, i+(n-j)-1] \neq T[j,\dots, n-1]$; otherwise, $LCE(i,j)=n-j$). Let $\textit{off}$ be a multiplicative offset. Throughout the procedure, the invariant $\textit{off}=2^e\mod q$ will hold in state $\langle i,j,e,k\rangle$. We start with $\textit{off}=1$. We use a modified version of Equation \ref{eq: rk substring} by adding a parameter (exponential $exp$) to the Karp-Rabin hash function:
	\begin{equation}\label{eq: rk substring1}
		\rk'(T[i,\dots, j],exp) = \rk(T[0,\dots, j]) - \rk(T[0,\dots, i-1])\cdot exp \mod q
	\end{equation}
	Note that $\rk(T[i,\dots, j]) = \rk'(T[i,\dots, j],2^{j-i+1})$. At binary search step $\langle i,j,e, k\rangle$ we still have to compare the last $l = k-e$ characters of $T[i,\dots,i+k-1]$ and $T[j,\dots,j+k-1]$. We split each of these two substrings suffixes in a left part of length $l' = 2^{\lfloor\log_2(l/2)\rfloor}$ (i.e. the closest power of 2 smaller than or equal to $l/2$) and a right part of length $l-l'$. Note that value $2^{l'}\mod q = z_{\log_2l'} = z_{\lfloor\log_2(l/2)\rfloor}$ has been pre-computed, so we can compute and compare in constant time the two values
	$$
	\rk(T[i,\dots,i+e+l'-1]) = \rk'(T[i,\dots, i+e+l'-1],\textit{off}\cdot z_{\lfloor\log_2(l/2)\rfloor})
	$$
	and
	$$
	\rk(T[j,\dots,j+e+l'-1]) = \rk'(T[j,\dots, j+e+l'-1],\textit{off}\cdot z_{\lfloor\log_2(l/2)\rfloor})
	$$
	If the two values differ, then we leave $\textit{off}$ unchanged and recurse on $\langle i,j,e,e+l'\rangle$. If the two values are equal, then we set $\textit{off}\leftarrow \textit{off}\cdot 2^{l'}\mod q = \textit{off}\cdot z_{\lfloor\log_2(l/2)\rfloor} \mod q$ and recurse on $\langle i,j,e+l',k\rangle$. Since $l/4 < l'\leq l/2$, this binary search procedure terminates in $\bigO(\log n)$ steps, each taking constant time. As done in the previous section, we can perform an exponential search before the binary search in order to reduce the size of the binary search interval from $\bigO(n)$ to $\bigO(\llen)$. Note that with our sampling $z_i$ of powers of 2 modulo $q$ it is straightforward to implement each exponential search step in constant time. 
	
	In order to minimize space usage, we can choose uniformly a prime $q$ such that $\lceil\log_2q\rceil = w$ (for the prime number theorem, there are $\Theta(2^{w-1}/(w-1))$ such primes). It is easy to see that $P'$ and $D$ can be constructed in $\bigO(n)$ time, therefore we obtain:
	
	\begin{theorem}
		Our data structure takes $n\lceil \log_2\sigma\rceil + n/w + w\log_2 n$ bits of space and supports extraction of any length-$\xlen$ text substring and LCE queries w.h.p. in 
		$\bigO\left(\xlen\log\sigma/w\right)$ and $\bigO(\log\llen)$ worst-case time, respectively. The structure can be built in optimal $\bigO(n)$ time.
	\end{theorem}

	\section{Implementation}
	
	Our solutions are very practical to implement. The C++ implementation of our structure based on Mersenne primes---available at~\cite{rklce}---is extremely fast and memory-efficient, supporting (on average) $710\  ns$ LCE queries and $160\ ns$ single-character access queries on a prefix of length $n=3\cdot 10^9$ of the Human genome ($\Sigma=\{A,C,G,T\}$). The space taken by the structure in RAM was of $750000560$ Bytes, i.e. only $560$ Bytes on top of the optimal 2 bits per character required to store the text. As a comparison, accessing at random positions an array of the same size takes---on the same machine---$74\ ns$ per access query. Our implementation took $170\ s$ to build the structure on the input file. We ran the experiment on a \texttt{intel core i7} machine running Ubuntu linux version 15.10. We leave to an extended version of this work a comparison of our software with other available implementations of other solutions for the LCE problem.
	
	\section{Final Remarks}
		
	In this work we showed that $o(n)$-time LCE queries are possible in the same space of the text. We achieve this result by combining Karp-Rabin fingerprinting with properties of Mersenne prime numbers. If a general prime is used in the Karp-Rabin hash function, then our general technique can still be applied but it requires $n/w+w\log_2 n$ additional bits of space. Both our data structures \textit{replace} the text, in that they support fast extraction of any text substring. 
	With respect to the Monte Carlo solution presented in~\cite{bille2015longest}, our structures are smaller, faster to build, and support faster LCE queries. Our results are also extremely practical. The implementation of our data structure based on Mersenne primes takes the same space of the input text and supports LCE queries with only a 10x slowdown with respect to direct access operations in main memory. The structure is also extremely fast to build.
	
	An interesting extension of this work would be to devise a way of 
	employing general prime numbers in the Karp-Rabin hash function without using $n/w$ additional bits on top of the space of the plain text. Another interesting line of research would be that of using derandomization techniques such as the ones proposed in~\cite{bille2014time,bille2015longest} in order to get versions of our solutions that always return the correct LCE.

	\bibliographystyle{splncs}
	\bibliography{rk-lce.bib}
	\bibliographystyle{plain}
	
\end{document}